# A kinetic model for qualitative understanding and analysis of the effect of 'complete lockdown' imposed by India for controlling the COVID-19 disease spread by the SARS-CoV-2 virus

Raj Kishore[1], Prashant Kumar Jha[1], Shreeja Das[1], Dheeresh Agarwal[1], Tanmay Maloo[1], Hansraj Pegu[1], Devadatta Sahoo[2], Ankita Singhal[3], Kisor K. Sahu[1,4]*

1. School of Minerals, Metallurgical and Materials Engineering, Indian Institute of Technology Bhubaneswar, India 752050
2. NetTantra Technologies (India) Pvt. Ltd. Bhubaneswar, Odisha 751021, India
3. Indian Institute of Technology Roorkee, Uttarakhand, India – 247667
4. Virtual and Augmented Reality Center of Excellence (VARCoE), Indian Institute of Technology Bhubaneswar, India 752050

*Email: kisorsahu@iitbbs.ac.in

**Abstract**

The present ongoing global pandemic caused by SARS-CoV-2 virus is creating havoc across the world. The absence of any vaccine as well as any definitive drug to cure, has made the situation very grave. Therefore only few effective tools are available to contain the rapid pace of spread of this disease, named as COVID-19. On 24$^{th}$ March, 2020, the the Union Government of India made an announcement of unprecedented 'complete lockdown' of the entire country effective from the next day. No exercise of similar scale and magnitude has been ever undertaken anywhere on the globe in the history of entire mankind. This study aims to scientifically analyze the implications of this decision using a kinetic model covering more than 96% of Indian territory. This model was further constrained by large sets of realistic parameters pertinent to India in order to capture the ground realities prevailing in India, such as: (i) true state wise population density distribution, (ii) accurate state wise infection distribution for the zeroth day of simulation (20$^{th}$ March, 2020), (iii) realistic movements of average clusters, (iv) rich diversity in movements patterns across different states, (v) migration patterns across different geographies, (vi) different migration patterns for pre- and post-COVID-19 outbreak, (vii) Indian demographic data based on the 2011 census, (viii) World Health Organization (WHO) report on demography wise infection rate and (ix) incubation period as per WHO report. This model does not attempt to make a long-term prediction about the disease spread on a standalone basis; but to compare between two different scenarios (complete lockdown vs. no lockdown). In the framework of model assumptions, our model conclusively shows significant success of the lockdown in containing the disease within a tiny fraction of the population and in the absence of it, it would have led to a very grave situation.

**Introduction**

In December, 2019 some doctors at Wuhan city in Hubei province of the People's Republic of China (PRC) started noticing peculiar profiles of some pneumonia patients [1-2]. These patients did not respond to the standard medical treatment protocol for pneumonia and some of their conditions deteriorated rapidly despite the best efforts of the doctors and sometimes even leading to fatalities. Later it was found to be caused by a new virus, named as SARS-CoV-2 [3], because of its relation to the outbreak of severe acute respiratory syndrome (SARS) during the years 2002–04 primarily in PRC and to a lesser extent in Canada, Singapore, Vietnam etc. SARS was caused by SARS coronavirus (abbreviated as SARS-CoV or SARS-CoV-1). It is so named because of the arrangement of the spike proteins of the virus that is reminiscent of a 'corona'. Since the virus for the present outbreak belonged to the same family as that of SARS, so it was named as SARS-CoV-2 (the number '2' denotes the second member of the same virus family, i.e., they are genetically connected) [4-5].

The initial outbreak was unfortunately followed by the new-year as per Chinese lunar calendar, when one of the largest human migration on planet earth happens because of the festivity associated with it in the world's most populous nation. Since Wuhan city is one of the most prominent industrial hub in the world's second largest economy, the virus silently followed the human traffic routes and dispersed across the entire globe. There are two classes of prime factors that stands out about the 'success' of this virus that cause such a serious global pandemic unforeseen in the modern, post-internet connected world. The first class of factors are biological and closely connected to the evolution of the virus itself. Let us very briefly gloss over it since this will not be the focus of this article, but very important to justify some of the assumptions made in this study. The SARS-CoV-2 virus has RNA as a genetic material which is chemically less stable than DNA. The RNA therefore is more susceptible to mutation and there is a chance that it might become less serious because of some random mutations. However, the SARS-CoV-2 RNA, that has approximately thirty thousand base-pairs [6] has an error correcting mechanism in-built and partly responsible for its success. The other biological factor is that the spike proteins in SARS-CoV-2 strongly binds to the Angiotensin converting enzyme 2 (ACE2) receptors in human. The ACE2 receptor provides the access to the virus inside human cells, where it highjacks the cell machineries and forces them to make the virus parts. Therefore the effective strong binding with ACE2 receptor is another key to its success. The third factor in this class is clinical. The SARS-CoV-2 affect both the upper and lower respiratory track. At the first stage, when the upper respiratory track is infected, the persons only has very mild or no symptom and unsuspected carrier becomes highly dangerous because that person does not even realize that he is a carrier and spreading the virus to the larger population. When the infection goes down the respiratory track, the symptoms becomes serious at times often with undesirable consequences even leading to fatalities.

However the second class of factors responsible for the 'success' of the virus is connected to its accidental capacity to exploit the human migration pattern and this will be the central theme of this article. As we have already discussed that, at the early stage when the infection is largely limited to the upper respiratory track, the affected person mostly mistake its symptom as that of a mild flu and become contagious. In the absence of any clinical preventive mechanism (such as vaccine) or any effective drugs to cure the infected persons, containment of the disease through clinical interventions is still largely an unsolved puzzle. Therefore, the only possibility to contain the rapid spreading of the disease in communities is identifying and isolating such type of carriers by clinical diagnosis, which WHO referred to as "Test, test and test" [7]. Such type of strategy is effectively adopted by countries like South Korea and Singapore. However, for a large and highly populated country like India, there are operational, clinical, infrastructural and financial limitations towards adopting this kind of strategy at least at the very early stage. So the other option is to deny an easy route to the virus that it can thrive on. Therefore, the Union Government of India took an unprecedented step of announcing a country wide 'complete lockdown' for 21 days, starting from 25$^{th}$ March, 2020 for its entire population of roughly 1.35 billion [8-9]. Meaning that, during this period the entire population were asked to remain confined within their homes, or wherever they stayed at that point of time and all kinds of movements were largely prohibited except only for a tiny fraction responsible for providing essential services. No exercise of similar scale and magnitude has been ever undertaken anywhere on the globe in the history of entire mankind. This study aims to scientifically analyze the implications of this decision using a kinetic theory model (discussed later) covering more than 96% of Indian territory (for some territories, reliable data are not available and therefore not included in the analysis; discussed later).

It has already been discussed that the disease is very effectively spread by unsuspecting carrier at the initial stage of infection limited at the upper respiratory tract. At this stage the virus is spread by droplet transmission by sneezing and coughing of the carrier. Moreover, very tight and effective binding of the virus with the ACE2 receptor of human, makes a facile human-to-human transmission. Therefore, this can be effectively modelled by the following protocol: a carrier or infected person coming in close contact with a heathy person will transmit the disease and make him also infected. There are some examples, where the spread of this virus is simulated using the same protocol [10-12]. Although this is not completely accurate, because it will depend on many other extraneous factors, but as a first approximation this is a satisfactory model, particularly in the absence of precise quantification of these extraneous factors at this early stage of analysis. Once we make this assumptions, we can use a readily available model: the molecular motion of gas which is closely connected to the kinetic theory of gas. In this model, each person will be represented by an 'atom' with significant differences, which will be discussed later. When an 'infected' (say color-coded by red) person come in contact with a healthy person (color coded by green), the healthy persons also get infected. It is imperative to mention here that, there are no available data source (at least to us) that may provide us accurate trajectory of each individual in the country. On the top of that, it will be computationally too expensive to model minute-by-minute movements of 1.35 billion individuals. Therefore, the best possible way out is to use the method of statistical representation. It is important to appreciate that the main objective of this study is to understand the disease spread in a local community, which we will refer in this article as 'clusters'. These clusters are very crucial from the viewpoint of 'containment zone' and developing operational plans for the high risk areas. For simplicity, we will distribute the Indian population in 29918 clusters each representing a population of ~ 4.51 thousands (rational for the choice of these numbers are provided in the 'methods' section). Now, there are two competing ways to reflect the true population density across the country: (i) using different sized clusters (or 'atoms' of different radius using the kinetic theory analogy) or, (ii) using same cluster size but disperse them according to population density. In this article the second method has been opted because of the following reasons:

a. If different sized clusters are used, then size of each clusters would need to be tracked, which is computationally expensive.

b. Each time a cluster is moved from one place to other, its size has to be rescaled based on the population density difference between these two locations, which is again expensive computationally.

c. In the second method, only care need be taken about setting the initial distribution, which should reflect the true population distribution of India. And since this article simulates the effects only for 60 days, during which India's population distribution is not expected to change significantly, therefore we need no expensive recalibration.

Although, the Union Government of India announced 'complete lockdown' of the entire country only for 21 days, we did simulate it for 55 days for academic interest. Deciding this kind of policy measures require serious considerations about multitude of other issues concerning societal, economic, law and order, emergency services, security, etc. just to name a few. Clearly, none of them are under the purview of present study.

We will make one more very important assumption: even if one person is infected in a cluster, we will assume the entire cluster is infected. There are two primary reason for this assumption: (i) Because of the close proximity, the other members in the same community are more vulnerable and all the other members in that cluster will be termed as 'high risk category' and, (ii) An individual cluster is the smallest representation of the population (we have no sub-cluster resolution). We are aware that this will lead to some overestimation and therefore, it might be perceived that it represents kind of a 'worst case scenario'. Since the aim is to qualitatively understand the effect of 'complete lockdown' by simulation the scenarios with and without the imposition of lockdown, the use of exactly same cluster distribution for both the case make it a case for fair comparisons. The starting point of this simulation is the prevailing situation as on 20th March, 2020.

**Methods**

Before describing the model in details, let us discuss some of the key features of this model, what this model does not aim to achieve and a word of caution.

**Salient features of the model in the present study:**

- It accurately represents the state wise population density distribution
- It accurately represents the state wise infection distribution for the zeroth day (20$^{th}$ March, 2020)
- It attempts to realistically represent movements of average clusters

- It attempts to capture the rich diversity in movements patterns across different states as per OLA's 'Ease of moving index'
- It aims to capture the different migration patterns across different geographies
- It also captures the different migration patterns of pre- and post-COVID-19 outbreak
- Effect of age on the probability of getting infected has been captured based on World Health Organization (WHO) report
- The Indian demographic data for each state is considered based on the 2011 census
- Capturing the effect of the incubation period of five days as mentioned in WHO report

**What this model does NOT aim to achieve:**

- Make a long-term prediction about the disease spread on a standalone basis [but to compare between two different scenarios (complete lockdown vs. no lockdown)]
- Make a verifiable and quantifiable estimation of disease dynamics
- Capture the effects of clinical interventions
- Capture the effects of long distance travels by individual using long-haul air/rail routes
- Capture the effects of primitive, but very effective, measures such as increased frequent proper washing of hand
- Capture the effects of stray events

**A word of caution: This model is strictly for qualitative comparisons only based on the assumption made.**

In order to maintain computational complexity to a somewhat reasonable level, we decided to have at most 5000 clusters in the most populous state, Uttar Pradesh (UP). Therefore each clusters represent a population of ~ 4.51 thousands, which leads to a total number of 29918 human clusters for the country as depicted in Fig. 1(a). The grey color zone indicates the areas for which data are not available to us, therefore not included in this study. The size and number of infected clusters are very small, it is difficult to visualize, therefore in order to visualize some of these clusters, a portion of the state of Kerala is zoomed and shown as an inset of figure of Fig. 1(a). The state and union territory (UT) wise distribution of the total 29918 clusters are presented in table 1. These cluster have significant different properties when compared to the 'atoms' of a gas: (i) gas atoms do not overlap, but two human clusters can mingle intensely with each other, therefore, even complete 'overlaps' are allowed, (ii) atoms conserves momentum during collision, whereas no such thing is applicable for hum clusters and finally, (iii) angular momentum is conserved for atomic systems (if explicitly modelled) which is again not applicable for human clusters. Needless to mention that the present simulations are confined in 2-dimensinal space bounded by the boundaries of Indian territories. As on 20$^{th}$ March (the zeroth day of simulation), there were 258 infected persons and according to our protocol, therefore, out of total 29918 clusters, we denoted 258 as infected clusters. The state and UT wise distribution of these infected clusters can be found in Table 1. At this point, one might argue that, because of initial concentrated distribution of infections, there could have been more than one infected persons in a given cluster. While that is true, we did not have the exact coordinates of each of these infected individuals, so we had no way to know the exact distribution. So while the state-wise distribution of both the parameters, the total population [13] as well as the total infected persons [14] are accurate to the best of our knowledge and available data, their distribution within a state are random. Therefore, it is imperative to realize that, this simulation will not reflect the accurate disease dynamics within the states. This is really not a serious concern at this point, since the goal of this article is NOT to make a long-term prediction about the disease spread but to scientifically analyze the implications of 'complete lockdown'. Therefore it is imperative to create an unbiased and representative distribution of both healthy population and infected persons, which has been achieved in this article. It is known [15] that an average Indian commuter transit ~35 kilometer daily, and that has been taken as a reference point about the daily movement of each healthy cluster in the absence of any other better estimate. While this is a tad overestimation for average Indian, but not a serious issue since we are simulating 'a worst case scenario' as described earlier. Moreover, the daily movement pattern was assumed to follow a normal distribution with the standard deviation values, which is based on the 'Ease of moving index'. This index is published by OLA [16] for the capital cities of each states, (with few exceptions, for which the data are not available, the national average was used), which was used as a representative for that state. Again, this is not an accurate representation for the entire state, but this is the best representation than can be achieved, rather than using trivially the same number (35 kilometer) for all the clusters for all the states, moreover at the initial phase, the disease spread is city (state capital) centric, therefore it gives some credence to this argument. This OLA ratings for the different cities of the Indian states are based on the local human behavior, transportation infrastructure, traffic rules, etc. in the corresponding cities therefore they capture a rich data diversity. Some fraction of human clusters (for example the migrant laborers) in a state are prone to migrate. Thus, in order to model this inter-state movement, we have considered the migration data available at [17] of each state and UT. The symptoms of this disease does not appear on the same day when the virus enters in the human body but it appears after few days which is called incubation period. We have used the incubation period of five days as mentioned in WHO report [18]. The pattern of the movements of migrant populations are different in pre- and post-COVID-19 outbreak and this has also been taken in account as depicted in Fig. 1(b). While, during the pre-COVID-19 the migration pattern follows the normal pattern of migration, during COVID-19 period, the migration patterns are reversed, to reflect the returning of migrant population to home. While, during incubation period, the movement of both heathy and infected clusters are identical, the movement of infected clusters, post the incubation period is reduced by 50 times that of normal movement, in order to capture the reduced movement once the disease (or the symptoms of the disease) are identified. During lockdown, the movement of all the clusters (healthy as well as infected clusters) are reduced with the same factor. We have also considered the effect of age on the probability of getting infected. This probability data is based on World Health Organization (WHO) report which published the demographic characteristic of COVID-19 outbreak in China during 16-24 February 2020 [18]. The Indian demographic data for each state is taken from the census 2011 [19].

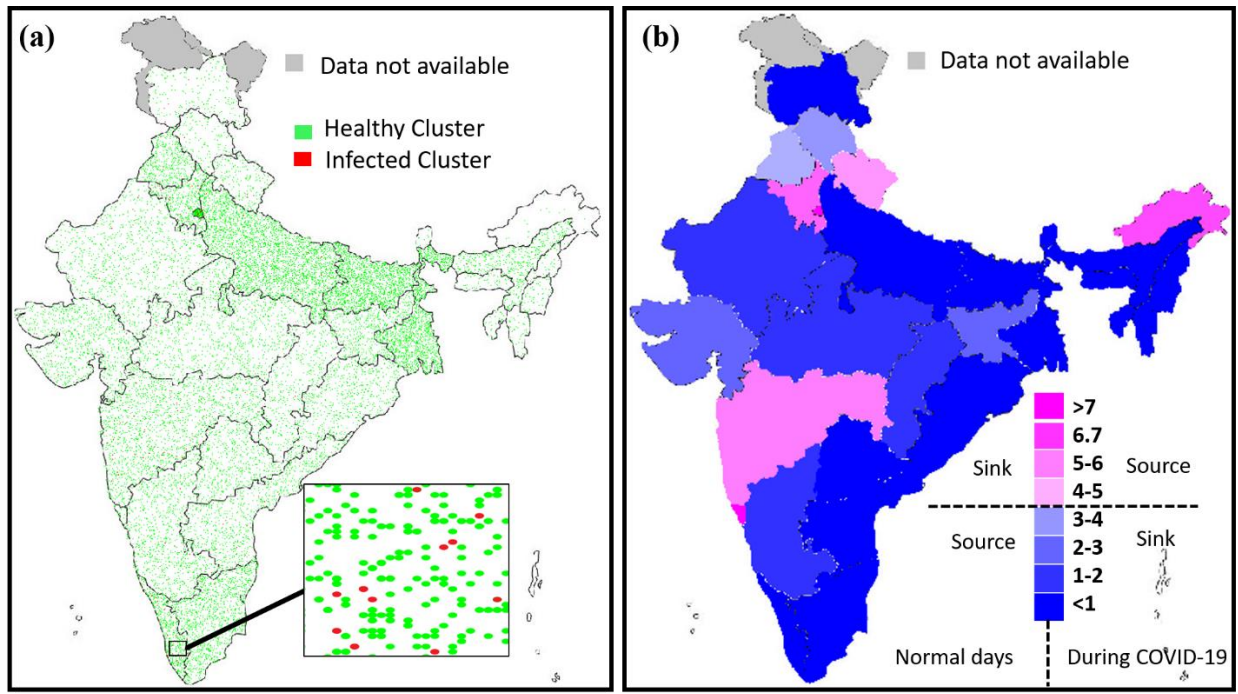

**Fig. 1:** The distribution of 29918 human cluster on the Indian map. The number of human clusters in each states are provided in Table 1 and calculated using the population density of the corresponding state. There are total 258 infected and 29660 healthy clusters, which are represented in red and green color respectively (see inset figure). The number of infected cluster in each state is taken based on the number of registered COVID-19 patients till 20[th] March 2020 in India [14], which is also the zeroth day of simulation.

**Table 1:** The number of human cluster and infected clusters (as on 20[th] March, 2020) considered in each state and UT at the   start of the simulation.

| State/UT | Human clusters | Infected clusters | State/UT | Human clusters | Infected clusters |
|---|---|---|---|---|---|
| Kerala | 852 | 40 | Uttar Pradesh | 5000 | 26 |
| Goa | 45 | 0 | New Delhi | 435 | 22 |
| Gujarat | 1411 | 7 | Himachal Pradesh | 167 | 2 |
| Rajasthan | 1679 | 19 | Uttarakhand | 266 | 3 |
| Punjab | 697 | 3 | Odisha | 1012 | 2 |
| Tamil Nadu | 1886 | 3 | Jharkhand | 746 | 0 |
| Karnataka | 1515 | 15 | Bihar | 2519 | 0 |
| Maharashtra | 2845 | 54 | Ladhak | 29 | 10 |
| Madhya Pradesh | 1755 | 4 | Jammu & Kashmir | 284 | 4 |
| Haryana | 619 | 19 | West Bengal | 2225 | 2 |
| Andhra Pradesh | 853 | 3 | Arunachal Pradesh | 36 | 0 |
| Telangana | 1294 | 19 | Assam | 773 | 0 |
| Chhattisgarh | 600 | 1 | Sikkim+Manipur+Nagaland+Tripura+Meghalaya+Mizoram | 375 | 0 |

The flow chart of the simulation is depicted in figure 2. The complexity of the simulation is of the order of $O(N^2)$, where $N$ is the number of human clusters considered. Although the complexity can be reduced by utilization of clever computational algorithms, but the main focus of this work was to make quickest implementation, so deliberate attempts were made not to use any such tricks to minimize unwarranted mistakes. The simulation is mainly divided into three steps:

(i) *Initialization*: In the initialization step, while the numbers of clusters inside a state truly represents its area and density distribution the spatial distribution of the clusters within a state is assigned randomly. The number of infected cluster ID is also assigned randomly based on the number of active cases in the corresponding state or UT.

(ii) *Selection of movement distance*: The cluster will move either a distance $M_h$ (distance travelled by a healthy cluster) or $M_i$ (distance travelled by an infected cluster). The selection of this distance, depends on mainly three criteria:
(a) Whether the lockdown is imposed or not?
(b) Whether the cluster is infected or healthy?
(c) Whether the incubation period (after which symptoms appears) of infected cluster is over or not?

(iii) *Neighbor detection*: After moving the cluster, its neighbor is detected based on the Euclidian distance between their geometrical centers. If the distance is less than the

*2.01\*R*, where, *R* is the radius of the cluster (assuming a 5% uncertainty factor in distance estimation). The value of *R* is 5.3 km, which is based upon pan India population density value.

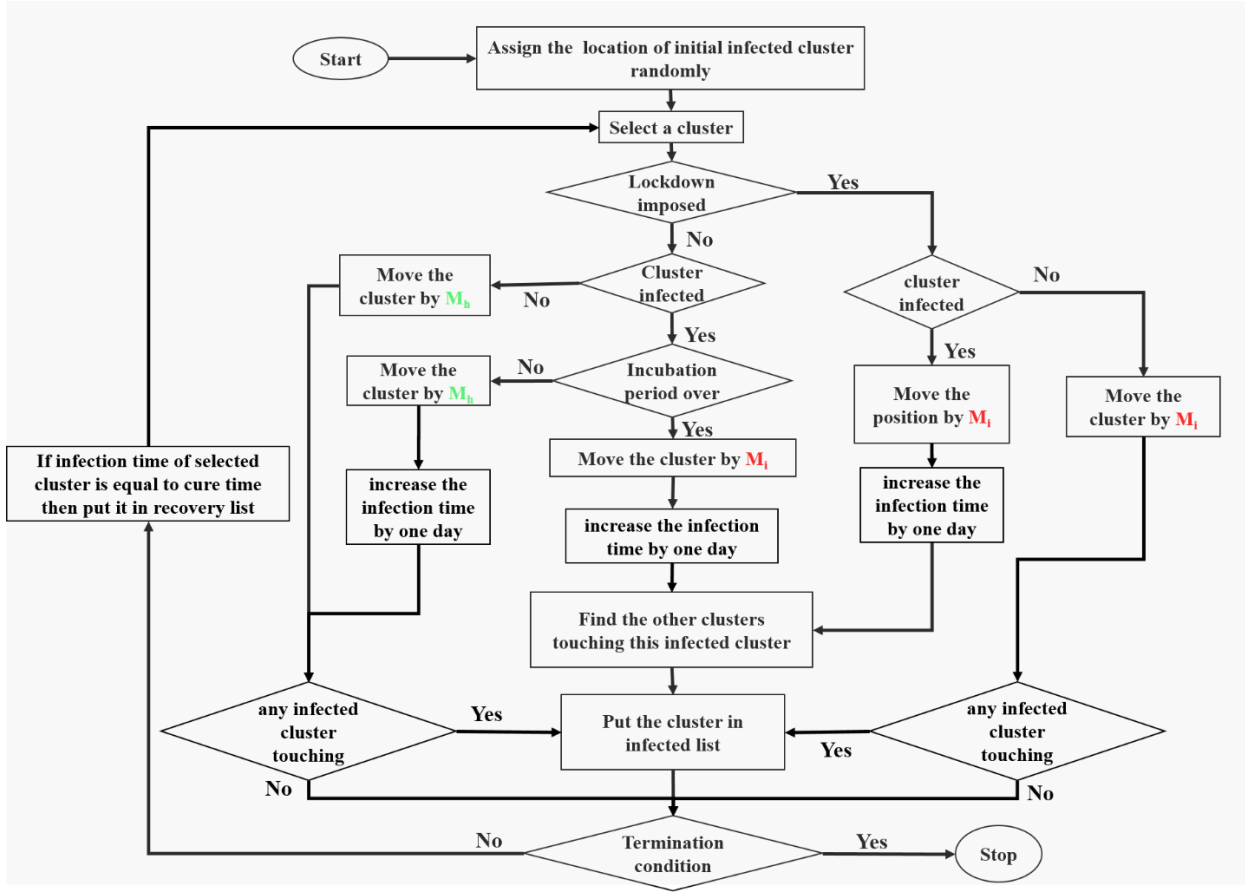

**Fig. 2:** Flow chart of the algorithm used in the COVID-19 simulator. The distance $M_h$ (movement of healthy cluster) and $M_i$ (movement of infected cluster) distribution is shown in supplementary figure S1_A and S1_B.

We have calculated the radius of the cluster, *R* as

$$R = \sqrt{\frac{1}{\pi}\left(\frac{P}{C} \times \frac{A}{P}\right)} \tag{1}$$

Where, *P* and *A* are the total population and area (in km²) of India respectively. *C* is the total number of clusters considered (29918 in the present study). The *P/C* ratio gives the number of people represented by one cluster (~4.51 thousands people) whereas the *A/P* ratio gives the area (in km²) covered by a single person. Thus the multiplication of these two ratios gives the area covered by each cluster. The distance between two clusters (*D*) is the Euclidean distance between their centers (Fig. 3). If any one of the infected clusters come in contact to a healthy cluster and the distance *D*, between them are led than 2.01\**R*, then both the clusters will be labeled as infected cluster.

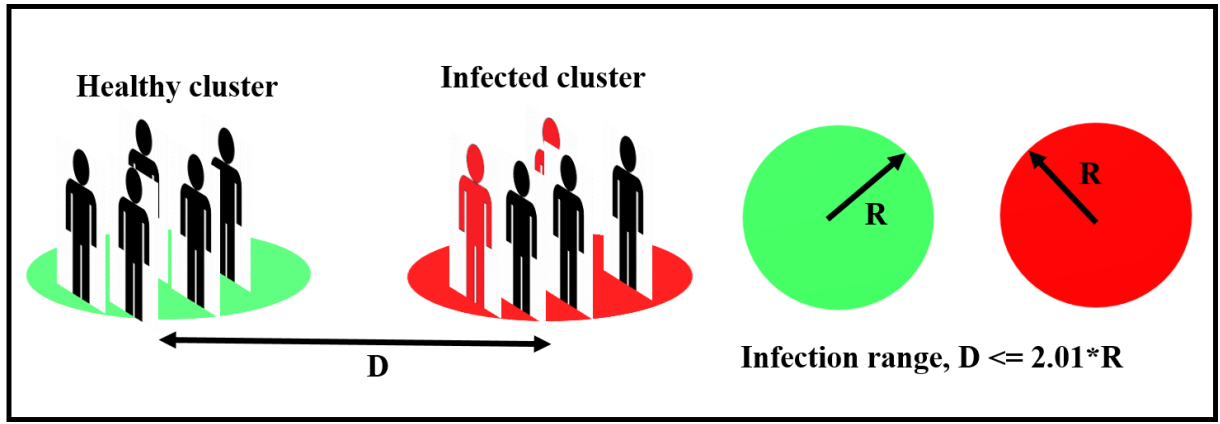

Fig. 3: The pictorial representation of healthy and infected cluster. Each cluster represent ~45100 people (obtained by dividing the total Indian population by number of clusters considered here). If a single person inside the cluster is infected, then we considered the entire cluster as infected cluster. The radius of both the clusters is R calculated by using Eq. (1).

**Results**

The spread of the infected cluster in the entire country for the next sixty days (from the zeroth day: 20$^{th}$ March) are simulated using the kinetic model as detailed in the method section for two different scenarios: (i) under complete lockdown on 5$^{th}$ day (25$^{th}$ March; when a nationwide 'complete lockdown' starts under the order of the Union Government of India) from the start of the simulation and (ii) without lockdown throughout the simulation. The result of 5$^{th}$, 15$^{th}$, 25$^{th}$, 35$^{th}$, 45$^{th}$ and, 60$^{th}$ day spread of COVID-19 under these two different scenarios are shown in Fig. 5, 6 and, 7. The effect of lockdown is clearly visible. There are very few new cases reported after lockdown imposed. Whereas, in other scenario, where the lockdown is not imposed, the disease outbreaks is rampant and almost entire map is filled with large number red (infected) clusters. Though in the second scenario also, the movement of infected clusters are restricted but the real cause of the disease outbreak is the frequent movement of the healthy as well as of those infected clusters in which symptoms are not visible (asymptotic during the incubation period). Whereas, during lockdown, the movement of all the clusters are restricted, thus preventing the outbreak of the disease on a far greater scale. Therefore, our analysis clearly presents this (identifying infection) as a great challenge and contact tracing at the earliest stage is the best possible strategy and the only available option at this stage. This, will significantly alter the disease spread dynamics and to a large extent will determine the final outcome. In the Fig. 4, the average movement of each cluster drops drastically from ~34 km per day to ~0.73 km per day (a drop of ~46 times) after lockdown imposed (at 5$^{th}$ day), whereas the movement in the scenario without lockdown is decreasing very slowly. The reason of the slowdown is not voluntary: as the clusters becomes progressively infected, they keep transmitting the infection during incubation, but post-incubations, the symptoms become prominent and the worried population restrict the movement (this has been incorporated in the modelling strategy and discussed in the 'methods' section). The effect of the drastic drop in the average movement in lockdown scenario is clearly visible in Fig.

8, where the number of active cases becomes almost constant after 14[th] day of the simulation (9[th] day from the lockdown).

There are few additional stray incidents that are not considered in the present simulation primarily because of lack of precise data. Although during the post-lockdown period large scale unnecessary movements were banned, there are several cases in which people denied/disobeyed the lockdown and kept continuously moving. Some of the prime examples are: (i) The construction and migrant workers from New Delhi and Mumbai started walking/moving by all sorts of available means to reach their homes mainly in Uttar Pradesh and Bihar) [20] (ii) There are several other cases also where people gathered and did not followed the social distancing rules. (iii) Many other stray incidents of not following the directives. It is pertinent to note that long distance travels by individual using long-haul air/rail routes were not modelled in this study. Therefore, the results presented in this article is not expected to represent the exact disease dynamics, but it provides an immensely useful tool to compare, assess and analyze the two scenarios (with and without lockdown) within the limits of assumptions made. Some of these factors are very important to determine the course of true dynamics, could not be captured in the present simulation for the lack of precise data at this stage and that is the primary reason for our assertion not to use this study as a long-term prediction on a standalone scenario. Additionally, few other factors have not been considered for the present study but can be easily implemented when the difficult time is over and data can be carefully mined for a more precise calculation such as: we have considered the uniform cluster distribution inside a state or UT based on its population density but in reality, most of the population of a state or UT lives in the Hi-Tech residential cities (e.g. Mumbai, Kolkata, New Delhi, Bengaluru, Pune, Bhubaneswar, etc.). The rest areas are either very sparsely populated or no one allowed to live there (like Forest, deserts, water lands, etc.). Same is true for the movement patterns. These can be taken care of at a later point of time for far higher accuracy. For example, though currently the Maharashtra has the highest registered positive COVID-19 cases (1135 till 8[th] April 2020 [8]), but ~95% of the cases are from the Mumbai, Pune and Thane which are either capital or economic cities of the state. This situation is nearly true for almost all states across the country, where most of the cases are from capitals and cities of economic importance. Thus for a more realistic kinetic model, the clusters should be distributed based on the district wise population density. The second main reason for this difference is the size of the cluster. Here, we have considered the same size of the cluster but in reality the size of these clusters should be very small, ideally speaking, one cluster should reflect one individual family. However sourcing the data for such a detailed model is a very serious undertaking both from the data mining point of view as well as from the required computational resources for such a simulation. However, if achieved, it will reflect the true disease dynamics both for highly populated regions like Hi-Tech cities (e.g. Mumbai, New Delhi, Kolkata, etc.) as well as for sparsely populated regions (e.g. Ladhak and Arunachal Pradesh, etc.).

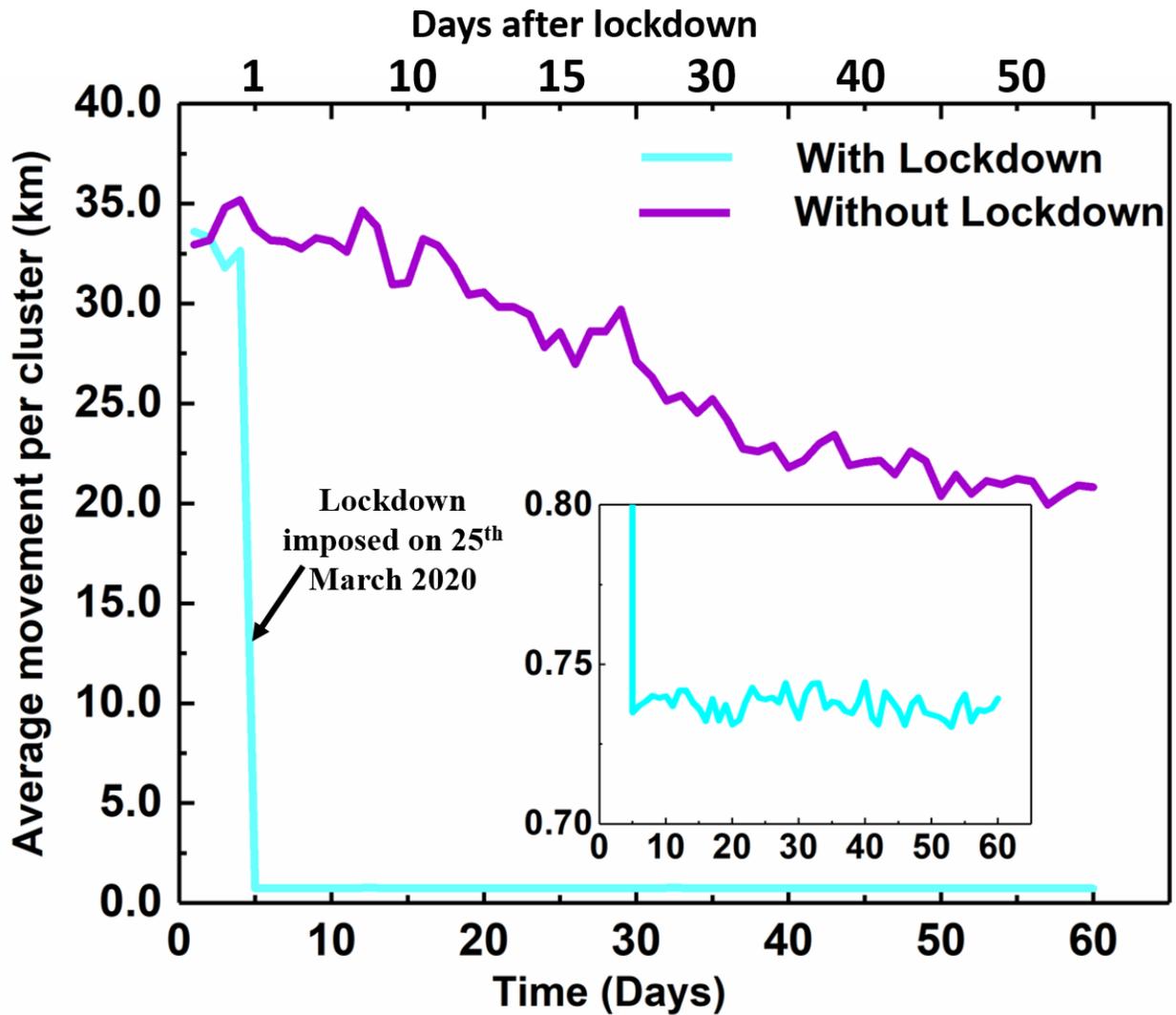

Fig. 4: The variation of the average movement per cluster in km with time in lockdown and without lockdown scenario.

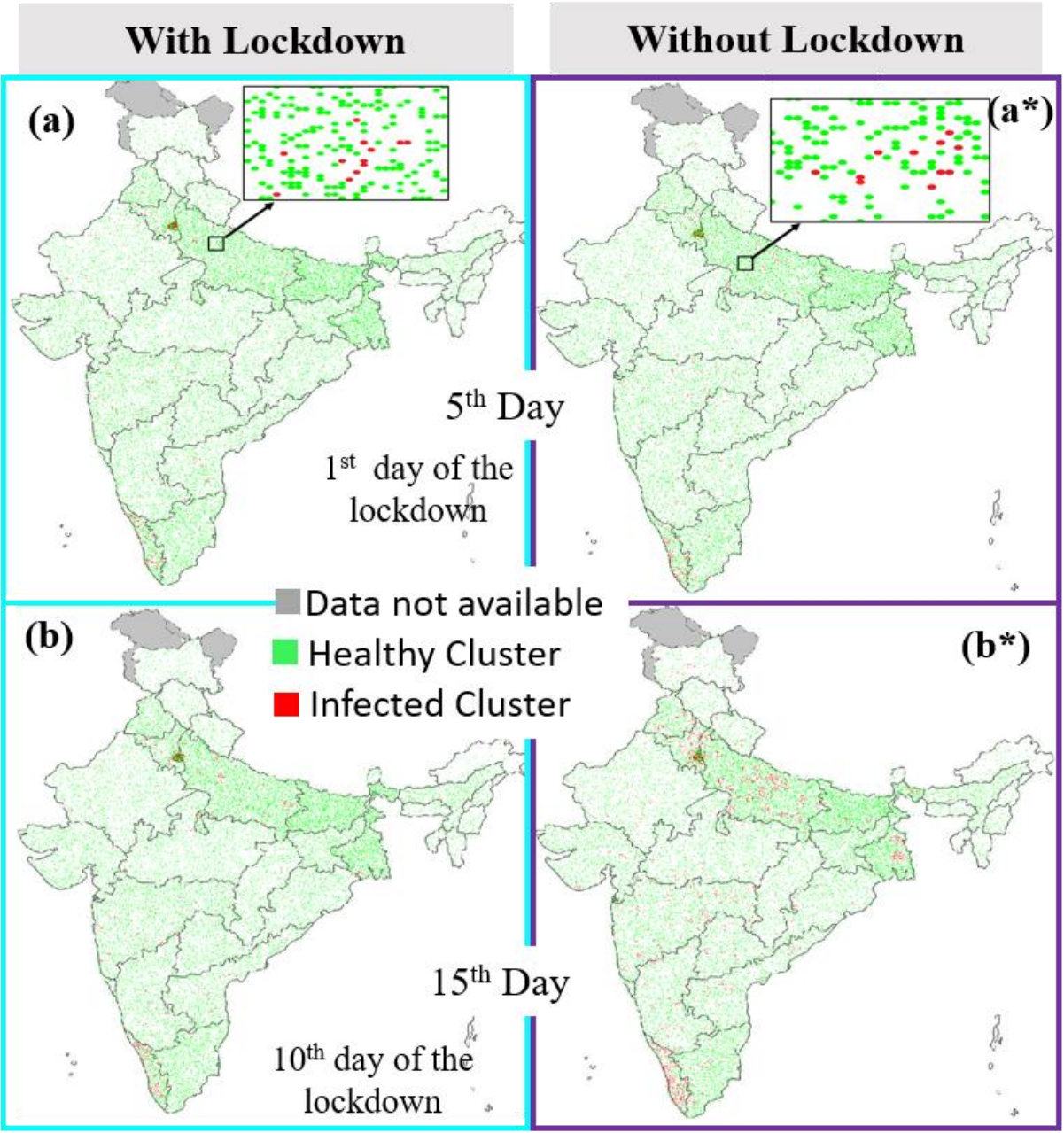

Fig. 5: The spread of COVID-19 clusters across the country on (a) 5$^{th}$ (b) 15$^{th}$ day, the lockdown imposed on 5$^{th}$ day onwards of the spread. The corresponding spread on (a*) 5$^{th}$ (b*) 15$^{th}$ day without lockdown in the entire duration

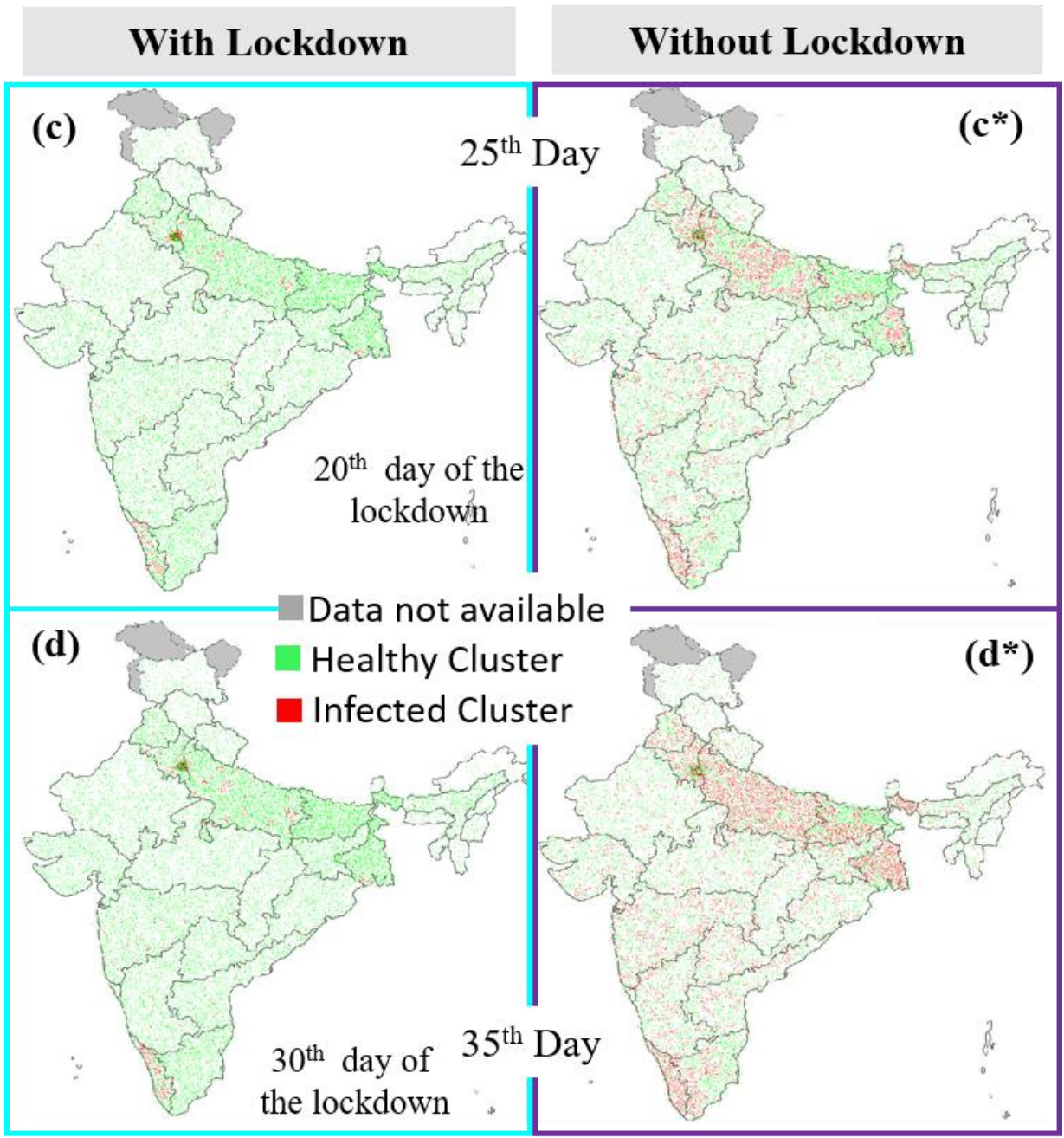

Fig. 6: The spread of COVID-19 clusters across the country on (c) 25$^{th}$ and (d) 35$^{th}$ day, the lockdown imposed on 5$^{th}$ day onwards of the spread. The corresponding spread on (c*) 25$^{th}$ and (d*) 35$^{th}$ day without lockdown in the entire duration

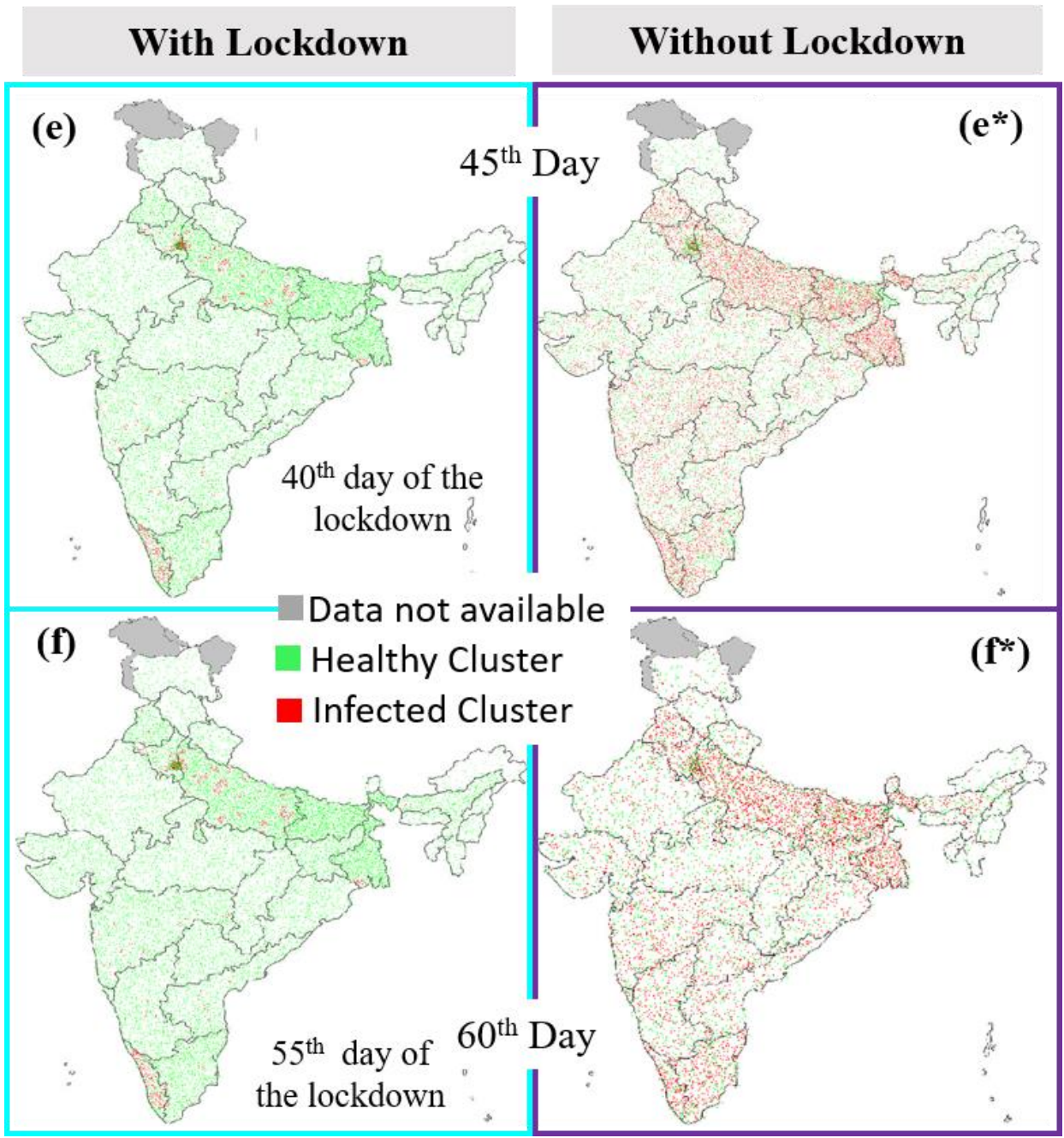

Fig. 7: The spread of COVID-19 clusters across the country on (e) 45th and (f) 60th day, the lockdown imposed on 5th day onwards of the spread. The corresponding spread on (e*) 45th and (f*) 60th day without lockdown in the entire duration

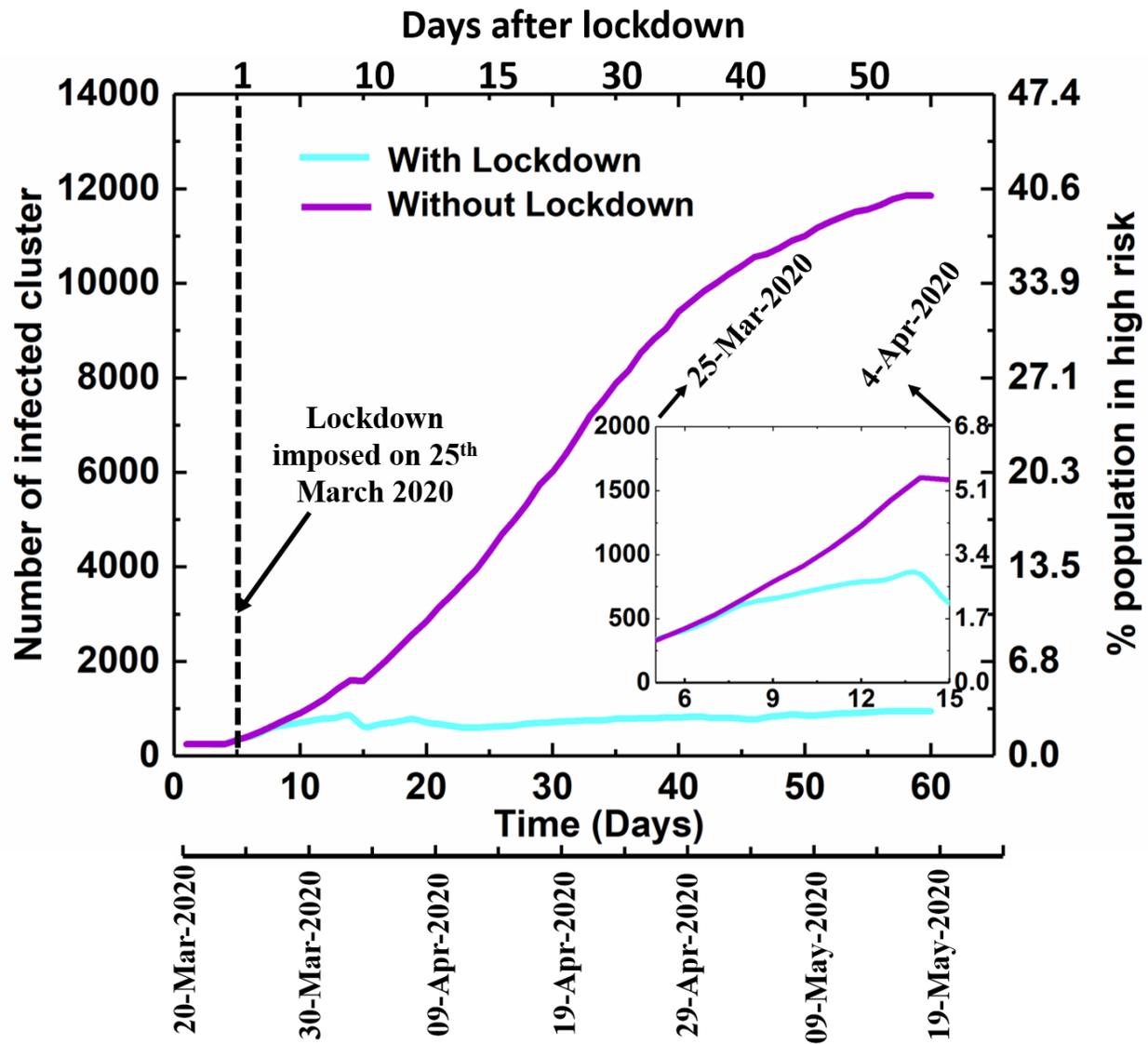

Fig. 8: The variation of the number of infected clusters with time in lockdown and without lockdown scenario. Secondary y axis represent the % of population in high risk category.

Figure 8 shows that, with complete lockdown in place, the infection could be restricted to ~1000 clusters while, without the lockdown it can be as high as ~12000 cluster after elapse of 60 days from zeroth day (20[th] March) or 55 days since lock down. This is equivalent to saying that, with complete lockdown in place, only less the 4% of population will be exposed to the 'high risk'. Here, 'high risk' indicates that: if there is at least one infected member in a community, all the other members in that community are at very high risk of getting infected if proper care are not taken, which is very difficult anyway because of the invisibility of the virus by the naked eye and therefore it cannot be known a priori which areas are contaminated and need to be disinfected. Without the lockdown in place, as high as 40% of the population will be exposed to the high risk, and worse still, that is not even the peak. While, the first scenario ('complete lockdown' enforced) did show an early peak (Fig. 8) that means the containment of the disease is successful to a great extent. This is a no mean feat for combating the disease, when modern medical interventions such as vaccine and curative drugs are not available and we had to resort to only primitive, but very effective, measures such as increased frequent proper washing of hand, following the social distancing rules. It is therefore heartening to see, how a policy measure, the 'complete lockdown' can effectively save a big populous country in this asymmetric war. The results produced by the kinetic model presented in this article demonstrate, beyond any reasonable doubt, the great efficacy of the 'complete lockdown' towards containing the disease COVID-19 caused by the SARS-CoV-2 virus in India.

**Conclusion**

The effect of unprecedented policy measure adopted by the Union Government of India for 'complete lockdown' of the entire country to contain the spread of COVID-19 has been analyzed scientifically with the help of a kinetic model. Extreme care were taken to choose large number of parameters from multitude of sources to reflect the prevailing ground realities in India. It was aimed to accurately represent state wise information in a statistical sense. The primary purpose of the present study is to compare the likely spread of COVID-19 in two different scenarios: (i) with complete lockdown in place from $25^{th}$ March, 2020 as it happened and, (ii) without any lockdown imposed. The simulation were run for 60 days starting from $20^{th}$ March, 2020, considered as the zeroth day. The simulation results conclusively shows that, in the first scenario, only a tiny fraction of population (less than 4%) are exposed to high risk category while in the second scenario more than 40% of population are exposed to high risk category. The difference is significant and therefore our detailed simulations very conclusively support the great efficacy of 'complete lockdown' as an effective tool for keeping the COVID-19 infection within a tolerable limit. This is an extremely important policy success for fighting COVID-19 disease in the absence of any preventive measures, such as a vaccine or any definitive drug for curing it.

**References**

[1] Wu, J.T., Leung, K., Bushman, M., Kishore, N., Niehus, R., de Salazar, P.M., Cowling, B.J., Lipsitch, M. and Leung, G.M., 2020. Estimating clinical severity of COVID-19 from the transmission dynamics in Wuhan, China. Nature Medicine, pp.1-5.

[2] Li, Q., Guan, X., Wu, P., Wang, X., Zhou, L., Tong, Y., Ren, R., Leung, K.S., Lau, E.H., Wong, J.Y. and Xing, X., 2020. Early transmission dynamics in Wuhan, China, of novel coronavirus–infected pneumonia. New England Journal of Medicine.

[3] Naming the coronavirus disease (COVID-19) and the virus that causes it. Accessed: 30 March 2020. https://www.who.int/emergencies/diseases/novel-coronavirus-2019/technical-guidance/naming-the-coronavirus-disease-(covid-2019)-and-the-virus-that-causes-it

[4] Zou, L., Ruan, F., Huang, M., Liang, L., Huang, H., Hong, Z., Yu, J., Kang, M., Song, Y., Xia, J. and Guo, Q., 2020. SARS-CoV-2 viral load in upper respiratory specimens of infected patients. New England Journal of Medicine, 382(12), pp.1177-1179.

[5] Walls, A.C., Park, Y.J., Tortorici, M.A., Wall, A., McGuire, A.T. and Veesler, D., 2020. Structure, function, and antigenicity of the SARS-CoV-2 spike glycoprotein. Cell.

[6] LeMieux J., Mining the SARS-CoV-2 Genome for Answers. Accessed: 30 March 2020. https://www.genengnews.com/news/mining-the-sars-cov-2-genome-for-answers/

[7] Atwoli L., Test, test, and test to effectively control Covid-19. Accessed: 30 March 2020. https://www.nation.co.ke/oped/opinion/Test--test--and-test-to-effectively-control-Covid-19/440808-5507216-3l2saxz/index.html

[8] Vaidyanathan G., People power: How India is attempting to slow the coronavirus, Nature April 2020. Accessed: 12 April 2020.

https://www.nature.com/articles/d41586-020-01058-5

[9] Kannan S., What India did -- and what it didn't -- in Covid-19 battle. Accessed: 8 April 2020.

https://www.indiatoday.in/news-analysis/story/what-india-did-and-what-it-didn-t-in-covid-19-battle-1663064-2020-04-03

[10] Stevens H., Why outbreaks like coronavirus spread exponentially, and how to "flatten the curve". March 2020. Accessed: 30 March 2020.

https://www.washingtonpost.com/graphics/2020/world/corona-simulator/

[11] Cleve Moler (2020). COVID19 Simulator. (https://www.mathworks.com/matlabcentral/fileexchange/74601-covid19-simulator), MATLAB Central File Exchange. Retrieved April 12, 2020.

[12] Chen, B., Shi, M., Ni, X., Ruan, L., Jiang, H., Yao, H., Wang, M., Song, Z., Zhou, Q. and Ge, T., Visual Data Analysis and Simulation Prediction for COVID-19. arXiv preprint arXiv:2002.07096.

[13] List of states and union territories of India by population. Accessed: 20 March 2020. https://en.wikipedia.org/wiki/List_of_states_and_union_territories_of_India_by_population

[14] India COVID-19 tracker. www.covid19india.org

[15] Gupta N. S., Indian commuters travel 35km/day says survey. Accessed: 20 March 2020.

https://timesofindia.indiatimes.com/business/india-business/indian-commuters-travel-35-km/day-says-survey/articleshow/63140954.cms?utm_source=contentofinterest&utm_medium=text&utm_campaign=cppst

[16] Ease of moving index-2018, OLA mobility institute. Accessed: 20 March 2020
https://ola.institute/ease-of-moving/

[17] Migration Tables, Census of India 2001. Accessed: 20 March 2020
http://censusindia.gov.in/Data_Products/Data_Highlights/Data_Highlights_link/data_highlights_D1D2D3.pdf

[18] Report of the WHO-China Joint Mission on Coronavirus Disease 2019 (COVID-19).
https://www.who.int/docs/default-source/coronaviruse/who-china-joint-mission-on-covid-19-final-report.pdf

[19] Population composition, Census of India. Accessed: 20[th] March 2020
http://censusindia.gov.in/vital_statistics/SRS_Report/9Chap%202%20-%202011.pdf

[20] Slater J., Masih N., In India, the world's biggest lockdown has forced migrants to walk hundreds of miles home. Accessed: 30 March 2020.

https://www.washingtonpost.com/world/asia_pacific/india-coronavirus-lockdown-migrant-workers/2020/03/27/a62df166-6f7d-11ea-a156-0048b62cdb51_story.html

# Supplementary Information

**S1:** The distribution of $M_h$ and $M_i$ for each cluster

The distribution of distance travelled by each cluster of a state/UT under unrestricted scenario is shown in Fig. S1_A and Fig. S1_B for all the states/UT considered in the present study. As mentioned in the "*method*" section, the distribution for all the states and UT is normal distribution with mean value at 35 km but the standard deviation for each state is different. The histogram plots for each states is also depicting the same distribution. Although the restricted movement distance $M_i$ is reduced by a factor of 50 but its distribution is again a normal distribution with mean distance value of 0.7 km and standard deviation for each state is same as used for $M_h$.

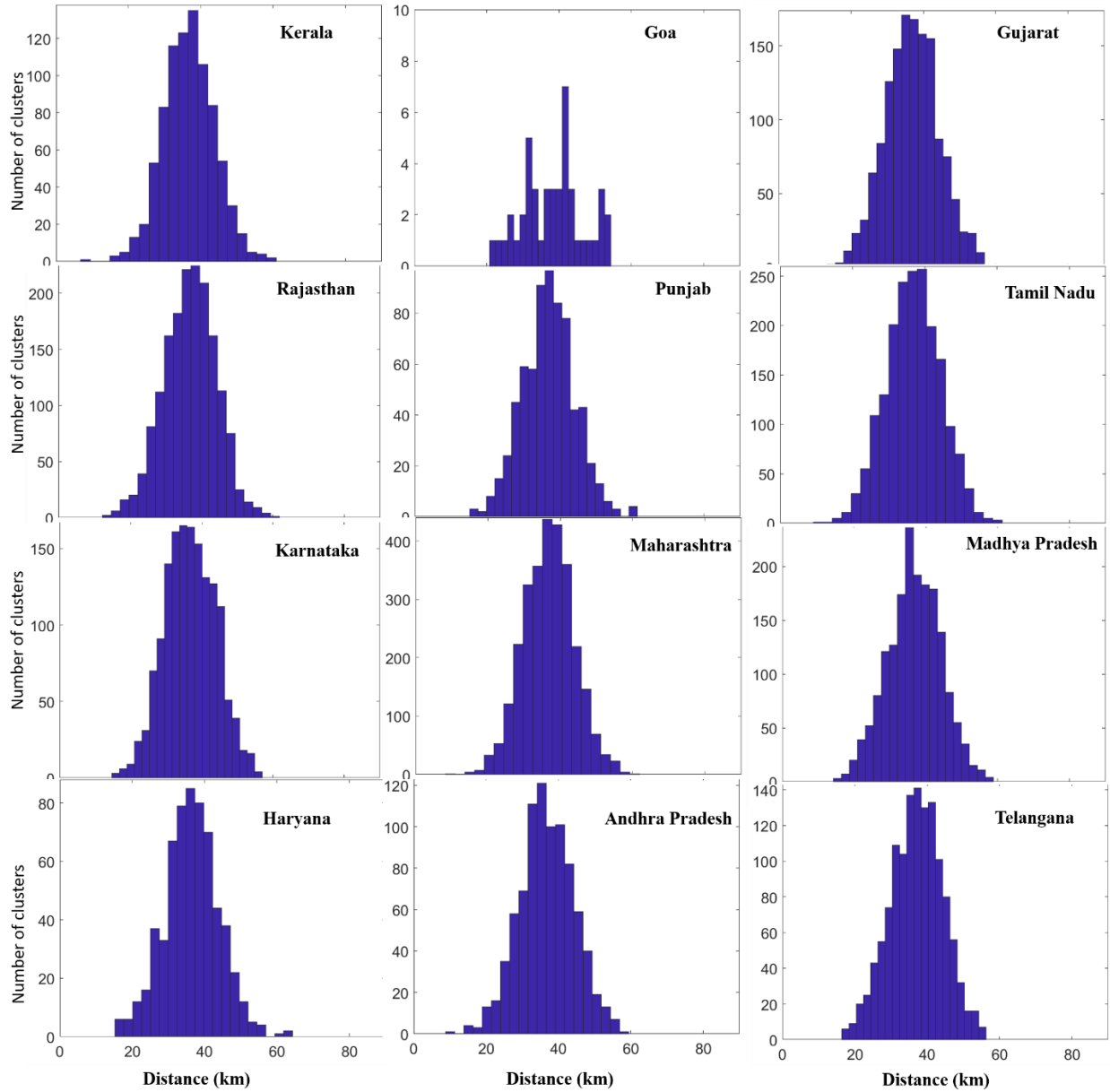

Fig. S1_A: The distribution of distance travelled by all the healthy clusters in unrestricted scenario, $M_h$ of Kerala, Goa, Gujrat, Rajasthan, Punjab, Tamil Nadu, Karnataka, Maharashtra, Madhya Pradesh, Haryana, Andhra Pradesh and Telangana (each panel is for one state/UT; name of state is mentioned at top-left corner of the panel).

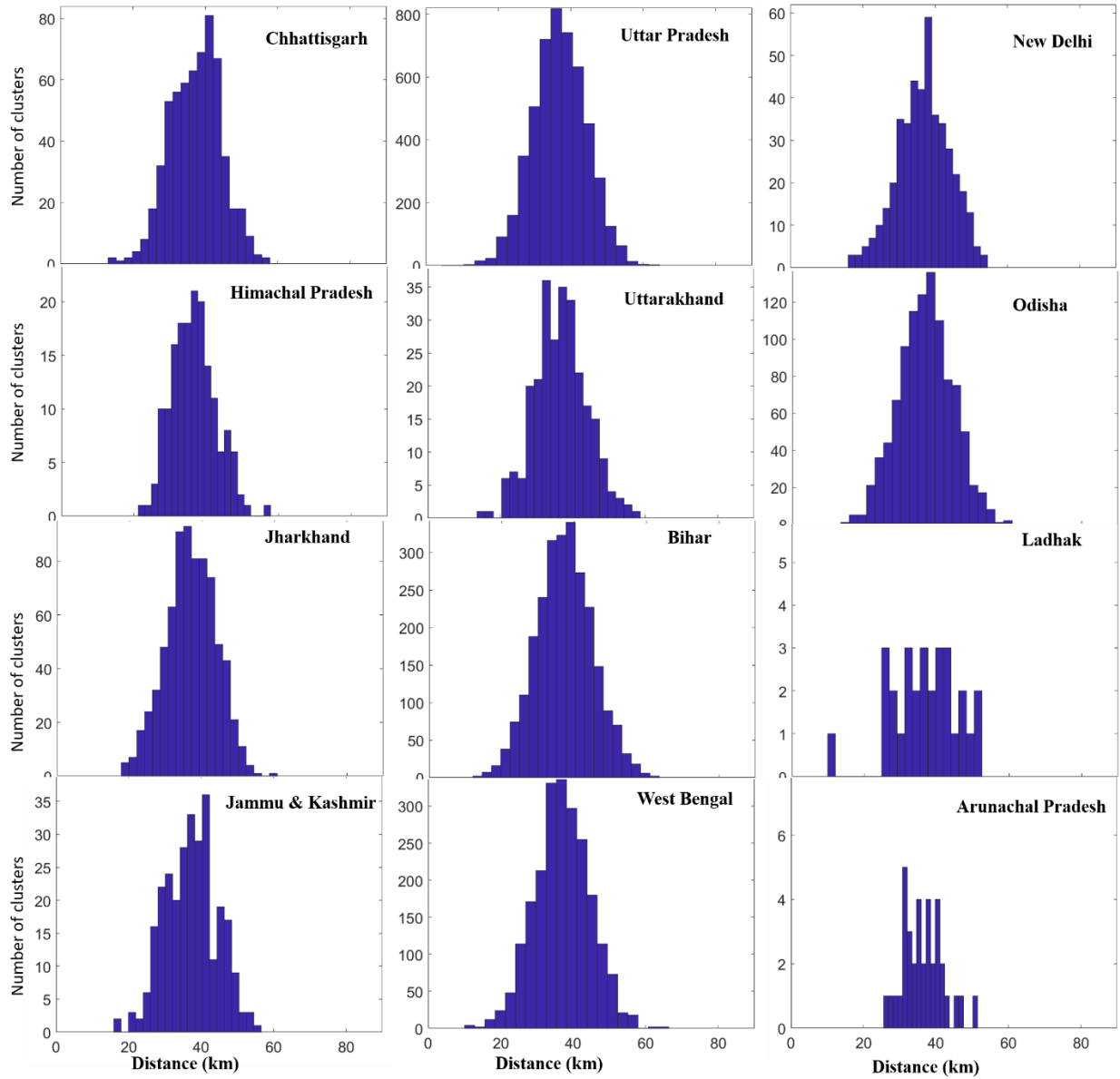

Fig. S1_B: The distribution of distance travelled by all the healthy clusters in unrestricted scenario, $M_h$ of, Chhattisgarh, Uttar Pradesh, New Delhi, Himachal Pradesh, Uttarakhand, Odisha Jharkhand, Bihar, Ladhak, Jammu and Kashmir, West Bengal and Arunachal Pradesh (each panel is for one state/UT; name of state is mentioned at top-left corner of the panel)